\documentclass[prl,
 reprint,
superscriptaddress,
 amsmath,amssymb,
 aps,
]{revtex4-1}

\usepackage{amsmath}
\usepackage{amssymb}
\usepackage{url}
\usepackage{color}

\usepackage{tikz}
\usepackage{pgfplots}
\usepgfplotslibrary{groupplots}
\usepackage[colorlinks=true,
           linkcolor=red,
            citecolor=blue,
            urlcolor=blue]{hyperref}

\usepackage{graphicx}
\newcounter{bla}

\begin{document}

    \title{The hot-tail runaway seed landscape during the thermal quench  in tokamaks}

    \author{Ida Svenningsson}
\affiliation{Department of Physics, Chalmers University of Technology,
  SE-41296 Gothenburg, Sweden}
\author{Ola Embreus}
 \affiliation{Department of Physics, Chalmers University of Technology,
   SE-41296 Gothenburg, Sweden}
 \author{Mathias Hoppe}
 \affiliation{Department of Physics, Chalmers University of Technology,
   SE-41296 Gothenburg, Sweden}
\author{Sarah L Newton}
 \affiliation{Culham Centre for Fusion Energy, Abingdon, Oxon OX14 3DB, UK}
 \author{T\"unde F\"ul\"op}
\affiliation{Department of Physics, Chalmers University of Technology,
  SE-41296 Gothenburg, Sweden}
    \begin{abstract}
Runaway electron populations seeded from the hot-tail generated by the
rapid cooling in plasma-terminating disruptions are a serious concern
for next-step tokamak devices such as ITER. Here, we present a
comprehensive treatment of the thermal quench, including the
superthermal electron dynamics, heat and particle transport, atomic
physics, and radial losses due to magnetic perturbations: processes
that are strongly linked and essential for the evaluation of the
runaway seed in disruptions mitigated by material injection. We
identify limits on the injected impurity density and magnetic
perturbation level for which the runaway seed current is acceptable
without excessive thermal energy being lost to the wall via particle
impact. The consistent modelling of generation and losses shows that
runaway beams tend to form near the edge of the plasma, where they
could be deconfined via external perturbations.
    \end{abstract}
    \maketitle    
One of the crucial problems facing magnetic fusion devices with large
plasma currents is the occurrence of plasma-terminating disruptions
\cite{Hender_2007}.  Such events can result in
a partial loss of magnetic confinement and a sudden cooling of the
plasma. This {\em
  thermal quench} (TQ) is associated with an increase in the plasma
resistivity and impedes current flow. As a result a strong inductive
electric field arises, which can lead to runaway acceleration of
electrons to relativistic energies \cite{Rosenbluth1997,Chiu_1998,Harvey2000,Boozer2015,Breizman_2019}.

Predictions show that a large part of the initial plasma current in
ITER, and other high-current tokamaks, can thus be
converted to a beam of energetic electrons
\cite{Boozer_2018,Sweeney2020}. The current carried by a ``seed'' of runaway electrons is greatly amplified in an avalanche process, with a multiplication factor exponentially sensitive to the initial plasma current \cite{Rosenbluth1997}. The subsequent
uncontrolled loss of these electrons could lead to localized power
deposition and damage of the plasma facing components (PFCs) \cite{Lehnen}.

The proposed disruption mitigation method in ITER is massive material
injection, using a combination of deuterium and impurities
\cite{HollmannDMS,Nardon_2020}. When injected in a pre-disruptive plasma, the
impurity atoms can radiate away the stored thermal and magnetic
energy, whilst the deuterium raises the critical electric field
required for runaway.  The mitigation system should be such that it
provides sufficient radiation to reduce conductive losses during the
TQ, to avoid damage on PFCs \cite{Lehnen}.

Material injection creates a two-component electron distribution,
consisting of hot electrons from the original bulk Maxwellian
population, and cold electrons ionized from the injected material
\cite{Chiu_1998,Harvey2000,Aleynikov2017}. The hot electrons lose energy due to the
interaction with the cold population, but in rapidly cooling plasmas,
due to the low collision frequency for fast electrons, a {\em
  hot-tail} often remains in the electron distribution \cite{Helander2004}.  Hot-tail generation is
efficient in the early phase of the disruption, the TQ, when the
temperature changes by orders of magnitude; in ITER, from tens of
kiloelectronvolts (keVs) to a few eVs in a millisecond (ms). This hot-tail
seed can be the dominant source of primary runaway electrons
\cite{Smith2005,Smith2008,Feher2011,Martin_Solis_2017}, yet remains
poorly modelled. The topology of the magnetic field confining the
particles also changes and stochastic regions are formed, allowing
rapid radial transport of runaway electrons  \cite{Izzo_2011, Sarkimaki_2020, Hu_2021}.  Therefore, a
large part of the hot-tail seed is expected to be deconfined.

Recent numerical studies indicate that assuming all hot-tail electrons
remain in the plasma overestimates the final runaway current in
experiments at ASDEX Upgrade
\cite{InsulanderBjork2020GO,Hoppe2021}. However, differences, for
example due to size and initial plasma current, may be expected
between disruption scenarios which can be studied on existing machines
and those which would occur in ITER.  Reliable predictions of the
hot-tail seed generation during the rapid TQ, in particular accounting for
the deconfinement of seed electrons, is needed to determine the
runaway conversion in disruptions.
Limitation of the runaway population by fluctuations during the subsequent current quench phase (dominated by avalanching) has been studied previously \cite{helander2000suppression, Svensson_2021}.
Self-consistent modelling of the
transport, power balance and runaway generation is essential also in the TQ, as these
effects influence the TQ dynamics and the associated induced electric
field, which in turn drives the runaway generation.

Here we present an integrated model of thermal quench dynamics,
including hot-tail generation and losses due to magnetic perturbations,
and use it to explore viable scenarios with combined deuterium and neon injection. We model the
current evolution together with the magnetic field fluctuation induced energy and particle
transport, as the injected material and bulk plasma evolve into a cold
free electron population and hot population, with densities
$n_{\rm cold}$ and $n_{\rm hot}$, respectively.

The total current density is given by the sum of Ohmic, hot electron and runaway  current densities:
    $
    j_\parallel = E\sigma_{\rm cold} + 
    \int_{|\boldsymbol{p}|<p_{\rm c}}ev_\parallel f \,\mathrm{d}\boldsymbol{p} + j_{\rm RE},
    $
where $\sigma_{\rm cold}$ is the Spitzer conductivity, $v_\parallel$ is the velocity parallel to the magnetic field, $f$ is
the superthermal (hot) electron distribution function, $p_{\rm c}$ the critical runaway momentum~\cite{Vallhagen2020} and the runaway current is defined as
  $
      j_{\rm RE} =  \int_{p_{\rm c} < |\boldsymbol{p}| < p_{\rm max}}ev_\parallel f \,\mathrm{d}\boldsymbol{p} + 
    ecn_{\rm RE}.
  $
Here, $n_{\rm RE}$ is the density of electrons having momentum $p>p_{\rm max}$ and
$E$ is the
electric field parallel to the magnetic field,
which in the cylindrical approximation evolves at radius $r$ according to 
\begin{equation}
\mu_0\frac{\partial j_\parallel}{\partial t} = \frac{1}{r}\frac{\partial}{\partial r}\left(r\frac{\partial E}{\partial r}\right).
\label{eq:ampere}
\end{equation}

The temperature of the cold electron population is determined from
the associated energy density
$W_{\rm cold} = (3/2) n_{\rm cold}T_{\rm cold}$, which is evolved according to
\begin{eqnarray}
  \frac{\partial W_{\rm cold}}{\partial t} &=\sigma_{\rm cold}E^2 - n_{\rm cold} \sum_i \sum_{j=0}^{Z_i-1} n_i^{(j)} L_{i}^{(j)}(T_{\rm cold},\,n_{\rm cold}) \nonumber     \\ 
  &   + Q_c + \frac{1}{r}\frac{\partial}{\partial r}\left[r D_W\frac{\partial T_{\rm cold}}{\partial r}\right],
  \label{eq:heateq}
  \end{eqnarray}
where $n_i^{(j)}$ is the number density of an ion with atomic
number $Z_i$ and charge number $Z_{0j} = j$, and $L_i^{(j)}$ the line radiation rates. The heat diffusion coefficient $D_W$ is calculated by integrating the kinetic radial diffusion coefficient $D$
over a Maxwellian corresponding to the cold electron population:
$D_W = n_{\rm cold}/(\pi^{3/2}m_ e^3v_{T}^3 T_{\rm cold})\int (m_e v^2/2)(v^2/v_T^2-3/2) D(\boldsymbol{v}) \exp(-v^2/v_{T}^2) \rm d\boldsymbol{p}$,
with $v_{T}=\sqrt{2T_{\rm cold}/m_e}$  the thermal velocity of the cold electron population. The rate of collisional energy transfer $Q_c = \int \Delta\dot{E}_{ee} f \,\mathrm{d}\boldsymbol{p} + \sum_i Q_{ei}$, where $\int  \Delta\dot{E}_{ee} f \,\mathrm{d}\boldsymbol{p}$ is the energy transfer from the hot electrons to cold free electrons, with $\Delta\dot{E}_{ee} = 4\pi n_\mathrm{cold} r_0^2 \ln\Lambda_{ee} m_e c^4/v$, $r_0$ the classical electron radius, and the sum is taken over ion species. The rate of collisional energy transfer between two Maxwellians is denoted $Q_{kl} = \left\langle nZ^2 \right\rangle_k \left\langle nZ^2 \right\rangle_l e^4 \ln\Lambda_{kl} [(2\pi)^{3/2}\epsilon_0^2 m_k m_l]^{-1} (T_k- T_l)\left({T_k}/{m_k}+{T_l}/{m_l}\right)^{-3/2}$, with $\left\langle nZ^2\right\rangle_k = \sum_{j=0}^{Z_k-1} Z_{0j}^2 n_k^{(j)}$ and the ion temperatures are evolved according to $(3/2)\partial (\left\langle nZ\right\rangle_ i T_i ) /\partial t = \sum_j Q_{ij}$.

The time evolution of the impurity species is described by $ \partial n_i^{(j)}/\partial t
= I_{i}^{(j-1)}n_i^{(j-1)}n_{\rm cold} -I_i^{(j)} n_i^{(j)}n_{\rm
  cold} + R_i^{(j+1)} n_i^{(j+1)}n_{\rm cold} -
R_i^{(j)}n_i^{(j)}n_{\rm cold}$, where $I_i^{(j)}$ include the
ionization rate and electron-impact ionization coefficients with the
 cold electrons, and $R_i^{(j)}$ are radiative recombination
rates, obtained from OpenADAS  \cite{ADAS}.  The number
density of cold electrons $n_{\rm cold}$ is such that the overall
plasma is charge neutral, satisfying $\sum_i \sum_j Z_{0j}n_i^{(j)}=
n_{\rm cold} + n_{\rm RE} + \int f \,{\rm d}\boldsymbol{p}$ at all
radii.

The superthermal electron
dynamics is determined by the kinetic equation, which in radius,
momentum and pitch coordinates ($r,\,p,\,\xi$), where $v_\parallel= v\xi$, reads
\begin{eqnarray}
  \frac{\partial f}{\partial t}& + eE\left(\frac{1}{p^2}\frac{\partial}{\partial p}\left[p^2\xi f\right]  +
    \frac{1}{p}\frac{\partial}{\partial \xi}\left[\left(1-\xi^2\right) f\right]\right) \nonumber\\
  &= \frac{1}{p^2}\frac{\partial}{\partial p}\left[p^3 \nu_s  f\right] +
  \frac{\nu_D}{2}\frac{\partial}{\partial \xi}\left[(1-\xi^2)\frac{\partial f}{\partial \xi}\right] \label{eq:kineq}\\
 & + \frac{1}{r}\frac{\partial}{\partial r}\left[ r D\frac{\partial f}{\partial r}\right],\nonumber 
  \end{eqnarray}
where we neglect the energy-diffusion term, an assumption strictly
valid only in the superthermal limit. Here $\nu_s(p)$ and $\nu_D(p)$ are
the slowing-down and deflection frequencies due to particle collisions
\cite{Hesslow2018}. 

The runaway density $n_{\rm RE}$
evolves according to
\begin{equation}
    \frac{\partial n_{\rm RE}}{\partial t} = 
    F_p(p_{\rm max})
    + \frac{1}{r}\frac{\partial}{\partial r}\left[r
      D\frac{\partial n_{\rm RE}}{\partial r}\right],
    \label{eq:runaway}
\end{equation}
where the flux from the superthermal region into the runaway region is $F_p = 2\pi p^2 \int (eE \xi  - p\nu_s) f \, {\rm d}\xi$, integrated along the upper boundary of the domain representing the kinetic hot electrons, $p_{\rm max}=3 m_e c$.
As we focus on the generation of the hot-tail seed during the initial phase of the disruption (the TQ)
we do not include 
here
the runaway growth due to avalanche multiplication, which occurs on a longer time-scale.

We have taken the Rechester-Rosenbluth form for the
coefficient $D = \pi q v_\parallel R \left(\delta
B/B\right)^2$ \cite{Rechester_1978} for simplicity, which assumes that the
magnetic field is fully stochastic. Here, $R$ is the major radius of the tokamak,  $\pi qR$ represents the parallel scale length of the magnetic perturbation, and $\delta B/B$
is the normalized magnetic fluctuation
amplitude, 
where in this work we assume the
fluctuating field amplitude to be constant in space and time. Although
such an assumption is not typical in the TQ, our results
provide a bound on
the perturbation level necessary to remove the hot-tail seed
entirely. The precise details of the transport coefficient
$D$ are relatively unimportant to the present study; we
present scans over its magnitude and are
mainly
concerned with its relative importance to the heat transport, and not
the electron evolution in any specific magnetic field configuration.

The plasma model
detailed here unifies
and extends components that have appeared in previous studies. Atomic screening effects~\cite{Hesslow2018} and fast-electron transport~\cite{Chiu_1998} extend the kinetic equation beyond that in~\cite{Aleynikov2017}.
The electric-field and current evolution has been widely used, e.g.~in Ref.~\cite{Martin_Solis_2017,Vallhagen2020}, but only recently in kinetic simulations~\cite{harvey2019time}. The thermal-energy equation is fully charge-state resolved, as
in the KPRAD model~\cite{KPRAD1997} or Ref.~\cite{Vallhagen2020}, and captures
electron heat transport from field perturbations~\cite{Rechester_1978} and non-thermal electron heating~\cite{Aleynikov2017}.

In the following, we  solve the coupled equations
   (\ref{eq:ampere})-(\ref{eq:runaway}) with the \textsc{Dream} code
(Disruption Runaway Electron Analysis Model), with full capabilities described in \cite{HoppeEmbreus}, and present results for an ITER-like disruption
with a combined deuterium and neon injection.
The initial plasma is fully ionized deuterium with a
pre-disruption electron density profile assumed to be uniform
with a value of $n_\text{hot}=10^{20}\,\rm m^{-3}$.
(Due to the limited appearance of the ion mass, the 50/50 deuterium/tritium composition  does not significantly alter the results.)
The initial
temperature profile of the electron and deuterium populations 
is given by $T(r) = 15 \left[1-(r/a)^4\right]
\,\rm keV$.  The initial current density profile is assumed to be
$j_\parallel(r)=j_{0}\left[1-(r/a)^4\right]^{3/2}$, with  $j_{0}$ chosen to give a total plasma current $I_p$. 
The major and minor radii are $R=6.2\;\rm m$ and $a=2\;\rm m$.
Neutral neon and deuterium are introduced with a prescribed density profile 
and zero temperature at the start of the simulation.

\begin{figure}
  \includegraphics{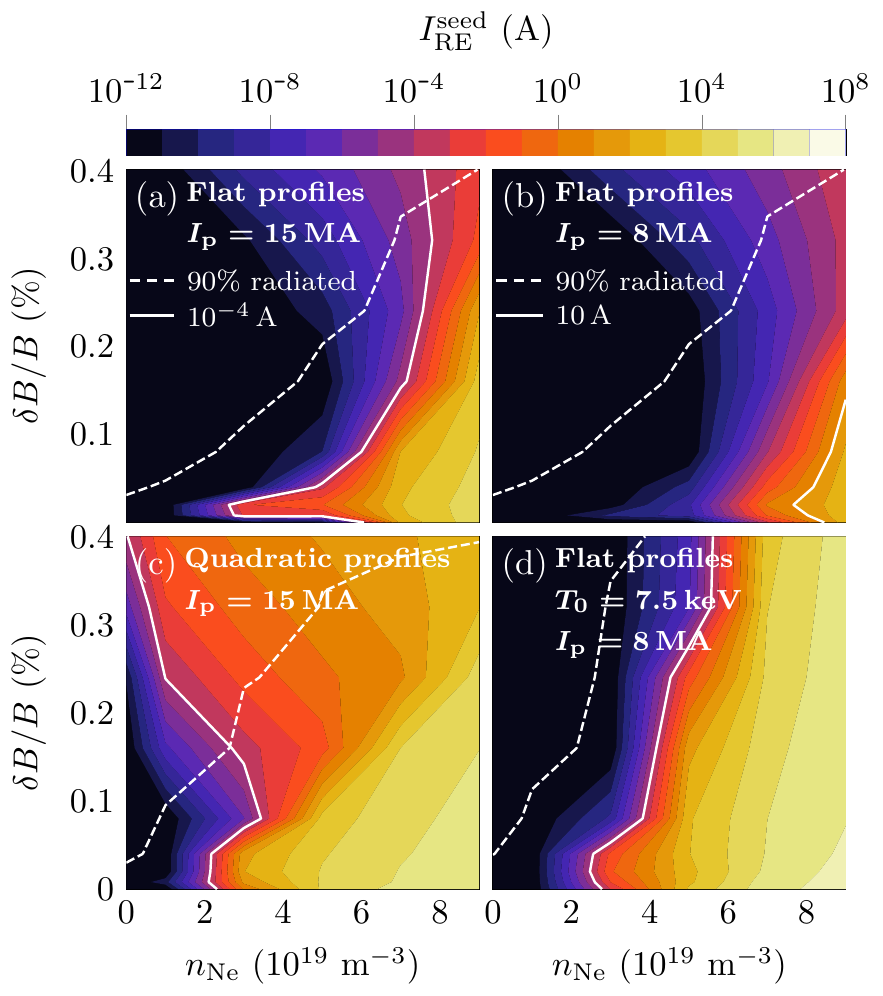}
\vspace{-5mm}  \caption{ Maximum seed runaway current as a function of injected neon density and normalized magnetic
    perturbation $\delta B/B$. The injected deuterium density is $n_D=10^{21} \;\rm m^{-3}$.
    Injected deuterium and neon profiles assumed to be (a,b,d) flat and (c) quadratic
    $n_{\rm Ne/D}(r) = n_{\rm Ne/D}\left[2/11+(18/11)(r/a)^2\right]$.
    Below the dashed lines the transport losses are acceptably low, and
    left of the solid lines the hot-tail seed is acceptably small. The initial central temperature is $T_0=15\;\rm keV$ in (a,b,c).}\vspace{-5mm} 
  \label{fig:contour}
\end{figure}

In order to assess the operational space leading to acceptable hot tail generation,  in Fig.~\ref{fig:contour} we present scans over injected impurity density and perturbation level, using the Rechester-Rosenbluth transport model   
with a radially constant magnetic
perturbation (taking $q\approx 1$). Each point in the plane corresponds to a TQ simulation,
with constant prescribed profiles of total impurity density 
and $\delta B/B$. Colors indicate the maximum runaway current reached during the 
simulation.
To avoid significant avalanche generation, the runaway seed current must be
lower than 10 A in the 8 MA case and $10^{-4}\,\rm A$ in the 15 MA
case \cite{Vallhagen2020}. These limits are indicated (for dominant hot-tail seed) with solid
lines.  To avoid damage to the first wall, at least 90\% of the
thermal energy loss must come from radiation \cite{Lehnen}. This gives
an upper limit on how strong the transport can be, which has been
indicated with a dashed line under the assumption that all kinetic
energy transported through the edge of the plasma will strike the
wall.

Figure \ref{fig:contour} shows that there is a region in parameter
space with moderate injected impurity density and radial transport
(between the solid and dashed lines), which gives acceptable hot-tail
generation and
non-radiative heat transport
to the first
wall. Simulations with only neon injection
(not shown)
indicate a very restricted acceptable operating space, due to a
lower radiated fraction.  For lower initial plasma current the
parameter region widens, as shown in Fig.~\ref{fig:contour}(b),
i.e.~greater impurity injection can be allowed, mainly due to the
increased acceptable seed currents.

Non-uniform impurity deposition profiles are likely to arise in realistic material injection scenarios. They can reduce the heat transported to the wall,
as radiation losses scale
quadratically with plasma density while transport scales only linearly. Therefore, in Fig.~\ref{fig:contour}(c) we investigate the effect of density profiles taking the form 
$n_{\rm Ne/D}(r) =
n_{\rm Ne/D}\left[2/11+(18/11)(r/a)^2\right]$, chosen so
the edge impurity density is ten times higher
than that at the
center, and the total number of ions corresponds to flat profiles at
the constant values $n_{\rm Ne}$ and $n_{\rm D}$.
This quadratic profile moderately widens the parameter region of tolerable transport losses (see the shift in the dashed line), however a significant increase in
runaway
current is observed, especially in the high-$\delta B/B$ region.
This increase of the runaway current due to transport is caused by the low density in the center of the plasma, allowing fast electrons to persist
due to the lower collisionality. When these core electrons are
transported to the cool edge -- where electric fields are stronger -- they can be accelerated as runaways. 

Figure~\ref{fig:contour}(d) illustrates the case with a lower initial
temperature, in which case hot-tail generation is more efficient (cf.~Fig.~\ref{fig:contour}(b)). This
counter-intuitive result agrees with earlier findings
\cite{Aleynikov2017}, and is due to the fact that in disruptions
dominated by radiation losses the TQ time
decreases faster with
temperature than the collision time. Thus, strong electric fields
capable of accelerating runaways are generated before the hot-tail
has had time to slow down.

\begin{figure}
  \includegraphics{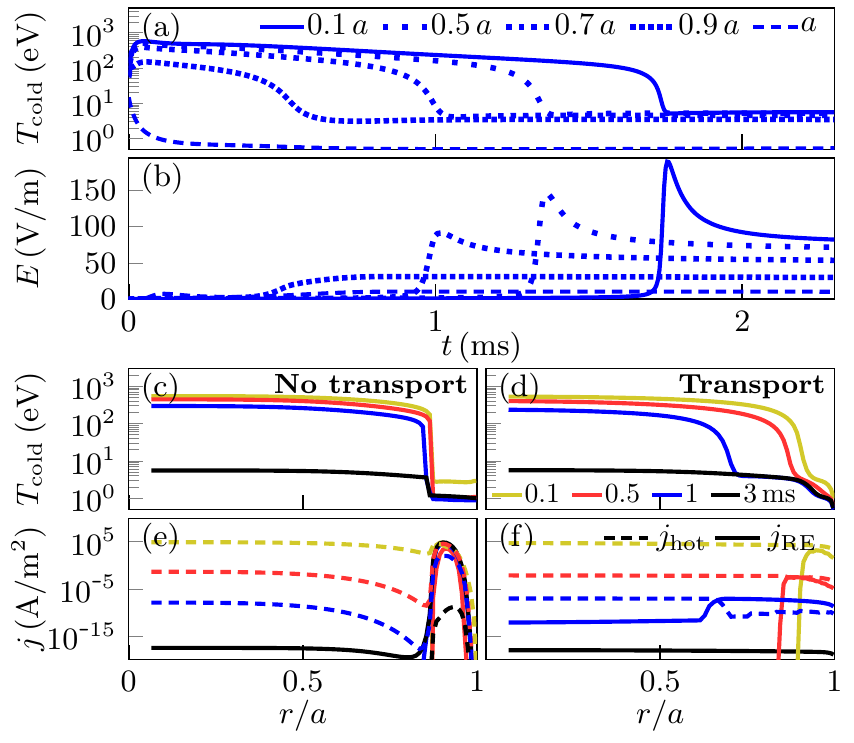}
\vspace{-5mm}  \caption{Simulation of an ITER-like disruption with initial plasma current $I_{\rm p}=15\;\rm MA$ and injected deuterium and neon densities  $n_\mathrm{D}=10^{21} \;\rm m^{-3}$ and  $n_\mathrm{Ne}=9\times 10^{19} \;\rm m^{-3}$, respectively. Injected deuterium and neon profiles assumed to be flat and in (a,b) $\delta B/B=0.16\,\%$.
    (a) Cold electron temperature as a function of time at five different radii.
    (b) Induced electric field at the same radial positions as in (a).
    (c,d) Cold electron temperature as a function of radius at four different times.
    (e,f) Hot (dashed) and runaway (solid) current density profiles at the same times as  in (c,d).
    (c,e) $\delta B/B=0$, (d,f) $\delta B/B=0.16\,\%$.
  }\vspace{-5mm} 
  \label{fig:temp}
\end{figure}

  To understand 
	in detail
	the basic dynamics of a TQ triggered by material injection, we show the evolution of temperature, electric field and current in  Fig.~\ref{fig:temp}.
The two upper panels show the cold electron temperature and induced electric field at different radial positions. In the two lower panels we compare the effect of radial transport on the radial evolution of the cold electron temperature and hot and runaway electron current densities. In Figs.~\ref{fig:temp}(c,e) no radial transport was assumed ($D=0$), and in Figs.~\ref{fig:temp}(d,f) $\delta B/B = 0.16\,\%$. According
to magnetohydrodynamic (MHD) simulations of disruptions induced by material injection in the
JET tokamak, the normalized perturbation amplitude during the TQ can
 be around $\delta B/B\simeq 1\%$ or higher
\cite{Sarkimaki_2020}.  The choice $\delta B/B = 0.16\,\%$ is a
conservative estimate, giving
a characteristic transport
time of runaways $t_{\rm d} = a^2 / D = a^2 / [\pi c R (\delta
B/B)^2] \approx 0.3\,$ms. The energy confinement time is a few ms, which is of the same order of magnitude as the
estimated TQ time in ITER \cite{Breizman_2019}.

The initial electron distribution function is a
Maxwellian, carrying
the initial density and temperature. The
injected material is ionized by the interaction with the hot plasma, cools the plasma due to line radiation, and
provides additional background electrons that are initially cold, but are heated via collisional energy exchange with the hot population, which 
quickly slows down as the plasma cools due to radiation.
At low temperatures Ohmic heating and radiation losses dominate,
and when in balance they support a stable equilibrium temperature typically in
the 5-20\,eV range.
In such regions the electric field tends to be strong, allowing efficient runaway acceleration and as a consequence radially localized current sheets may arise \cite{putvnjm,eriksson,Feher2011}. 
Near
the plasma edge, due to the lower current density and initial
temperature, the background electrons enter the cold equilibrium near
5\,eV at the beginning of the TQ. 
Meanwhile, in the core of the plasma,
Ohmic heating and energy transfer from hot electrons can
heat the background to hundreds of eV, which can be sustained for
multiple ms before the cold equilibrium is ultimately
reached.
A strong electric field is then induced, as illustrated in Fig.~\ref{fig:temp}(b). This occurs first at the edge, and
propagates inwards as the cold front reaches the central parts of the plasma. 

Figures~\ref{fig:temp}(e,f) show the radial evolution
of the plasma current. 
The initial current carried by hot electrons is rapidly converted to
Ohmic current carried by the cold background plasma as the 
hot electrons are slowed down by collisions.
The hot-tail electrons that are accelerated by the electric field and enter the
runaway region are shown with solid lines.
Radial transport of electrons in stochastic fields causes the central temperature to reach the cold equilibrium point at an earlier time, causing an earlier onset of strong electric fields and subsequent increased runaway acceleration. Note that the first runaway acceleration occurs at the outer radii, which might lead to hollow runaway density profiles. The reduction in the maximum runaway current density $j_{\rm RE}$ in Fig.~\ref{fig:temp}(f) compared to Fig.~\ref{fig:temp}(e) is due to the transport by magnetic perturbations. This indicates that inducing strong edge fluctuations can be useful in mitigating the runaway population produced by the strong edge cooling.

\begin{figure}
  \centering                 
  \includegraphics{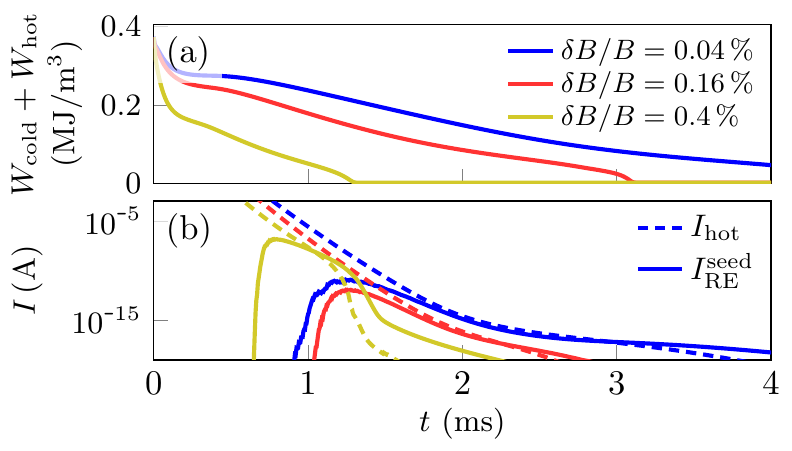}\vspace{-5mm}
  \caption{(a) Electron energy density  and (b) current for $\delta B/B=0.04\,\%$ (blue),  $\delta B/B=0.16\,\%$ (red) and  $\delta B/B=0.4\,\%$ (yellow) and injected densities $n_{\rm D}=10^{21} \;\rm m^{-3}$, $n_{\rm Ne}=5\times 10^{19} \;\rm m^{-3}$.}\vspace{-5mm}
  \label{fig:nonmonotonic}
\end{figure}

The simulations also show that the maximum runaway current
varies non-monotonically with magnetic perturbation $\delta B/B$. 
This is illustrated in Fig.~\ref{fig:nonmonotonic}, where we
show examples of the energy density $W_{\rm cold} + W_{\rm hot}$ and current
evolution for three different values of $\delta B/B$, where
$W_{\rm hot}=\int (\gamma-1)m_ec^2f\,\rm d\boldsymbol{p}$. 
With the two lower values, $\delta B/B=0.04\,\%$ (blue) and $0.16\,\%$ (red),
the runaway current (solid) peaks at similar
times, but since more hot
particles remain in the plasma
when $\delta B/B=0.04\,\%$,
this case results in
a higher peak value. The highest value $\delta B/B=0.4\,\%$ results in the
shortest TQ (yellow line in Fig.~\ref{fig:nonmonotonic}(a)), which causes
an earlier rise in the electric field
and earlier runaway acceleration. Previous predictions have suggested that 
the hot-tail seed would decrease monotonically with TQ duration~\cite{Smith2009}. Our
results indicate that the interplay between fast-electron transport and temperature evolution must be
considered
carefully in order to predict the seed current.

Reduced MHD simulations~\cite{Izzo_2011} indicate natural disruption activity gives fluctuations peaked in the core.
Such radially varying magnetic
perturbations give quantitative differences, reducing the acceptable parameter space, but the
hot-tail landscape is similar.
As expected, the runaway seed is larger
if the $\delta B/B$ radial profile is peaked at the core.
Optimising the net profile by externally applied perturbations, exploring transport coefficients motivated by the MHD simulations, provides a potential route to identify robust operation spaces.

In this work, we assumed instantaneous deposition of the injected
material, valid when the deposition time-scale is much shorter than
the time-scale of plasma cooling, which is often the case in pellet
injection experiments in medium-sized devices \cite{PazSoldan2020}.
When the time scales are comparable, which is the case for larger
devices such as ITER (traversing a 2 m radius in 2 ms requires a 1 km/s pellet speed, much higher than the few hundreds of m/s foreseen \cite{Baylor2015}), the model presented here should be complemented by an impurity injection model, such as pellet ablation.
The resulting non-uniform cooling as
the pellet
crosses the plasma has
implications for the limits on impurity density and perturbation level,
as we have illustrated with the non-uniform deposition profile. However,  in those cases where alarming runaway currents are
predicted, they tend to form near the edge of the plasma, where they can
more easily be deconfined by e.g.~external perturbations~\cite{Svensson_2021,SmithBoozerHelander} 
during the current quench. A potentially important effect not addressed here is the MHD stability of such current profiles, that might lead to safe termination \cite{Reux2021}.
 
In summary,
our results 
show that in the case of moderate impurity and
deuterium injection, radial transport caused by magnetic fluctuations
during the TQ should allow for efficient losses of hot-tail
seed
runaways in ITER-like disruptions, without producing excessive heat
loads on the first wall.  The cutoff value of impurity density below
which the runaway current is acceptable depends on the material
deposition profile, the pre-quench temperature, as well as the possibility of exciting magnetic perturbations.

\vspace{1cm}
The authors are grateful to E.~Nardon, I.~Pusztai, G.~Papp and the rest of the Plasma Theory group for fruitful discussions.  This work was supported by the European Research Council (ERC) under the European Union’s Horizon 2020 research and innovation programme (ERC-2014-CoG grant 647121) and the Swedish Research Council (Dnr.~2018-03911).

\bibliography{./ref}

\end{document}